\documentstyle[12pt,aps]{revtex}
\vspace{2cm}
\begin{document}
\title{\ \\ \ \\ \ \\ \ \\ \ \\  \ \\
COULOMB BLOCKING OF TUNNELING:\\ FROM ZERO-BIAS ANOMALY TO COULOMB GAP}
\author{L. S. Levitov}
\address{Massachusetts Institute of Technology,
12-112, 77 Massachusetts Ave., Cambridge, MA 02139, USA}
\author{A. V. Shytov}
\address{L. D. Landau Institute for Theoretical Physics, 2, Kosygin st.,
Moscow, 117334, Russia}
\maketitle
\thispagestyle{empty}
\bigskip

{\small
We present
an effective action approach for the problem of Coulomb blocking
of tunneling. The method is applied to the  ``strong  coupling''
problem  arising  near  zero  bias,  where  perturbation  theory
diverges. By a semiclassical argument, we obtain
electrodynamics  in  imaginary  time,  and  express  the anomaly
through exact conductivity of the system $\sigma(\omega, q)$ and
exact interaction. The calculation is tested by  comparing  with
the  known perturbation theory result for the diffusive anomaly.
Also, we use the method to study the anomaly enhancement due  to
external   magnetic  field,  and  the  effect  of  screening  by
electrodes.}
  \bigskip

\section{Introduction}
Suppression or enhancement of the tunneling conductivity near
zero bias is known to be a signature of interaction in the
system. It is called ``zero-bias anomaly'', and it has been
studied in metals and semiconductors
since the early sixties\cite{review}.
Initially, the origin of the anomaly was attributed to the Kondo
effect due to magnetic impurities. However, later it was
understood that a much more common mechanism is Coulomb blocking
of the tunneling.
A perturbation theory of this effect was developed by Altshuler,
Aronov and Lee\cite{AAL}. The theory deals  with  the  diffusive
limit, and shows that the blocking increases at small bias which
leads to a singularity in the tunneling conductivity. The theory
has  been  thoroughly  tested  experimentally  and  found  to be
extremely accurate when the anomaly is a weak feature on the top
of a large constant conductivity.

In the last years the interest shifted to the
systems with strong Coulomb effects, such as disordered metals
and semiconductors near metal-insulator transition\cite{Dynes}.
It is found that the Coulomb anomaly is sharply enhanced near
the transition, providing a test of Coulomb correlations.
Another important discovery is the observation by Ashoori et al.
of the Coulomb blocking of tunneling in a two-dimensional metal
in magnetic field\cite{Ashoori}. In this experiment it is found
that at certain magnetic field the zero-bias anomaly abruptly
increases and transforms to a ``soft Coulomb gap''. It has been
pointed out\cite{Eisenstein} that this transition is induced by
disorder. More recently, the gap was studied in the systems with
higher mobility\cite{Eisenstein,gap}, where current is almost
entirely blocked below certain threshold bias. These findings
caused a lot of theoretical work, concerned with the behavior of
current near zero bias\cite{HeEtAl}, and with determining the
gap width\cite{gap}. The anomaly has been shown to be
particularly interesting in the $\nu=1/2$ Quantum Hall
system\cite{HeEtAl} as a sensitive probe of the Quantum Hall
state.

Our goal in this contribution is to describe an effective action
theory\cite{LevitovShytov} that treats the anomaly as
cooperative tunneling. In the calculation, the tunneling rate is
related to the action for charge spreading and is expressed
through the actual conductivity $\sigma(\omega,q)$ of the
system. The treatment is non-perturbative, and remains accurate
in the strong coupling regime. For example, at the
metal-insulator transition, conductiviy has scale invariant form
$\sigma(\omega)\sim \omega^\alpha$, and we are able to predict
the form of the tunneling $I-V$ curve. This relation may be used
to determine the critical exponent $\alpha$. We discuss it
below, and also apply the method to diffusive anomaly in a two
dimensional metal, with and without magnetic field. Our results
are consistent with the perturbation theory
results\cite{AAL,HeEtAl}.

\section{Qualitative discussion}
Tunneling of an  electron  into  a  metal  involves  two  steps:
traversing  the  barrier, followed by spreading within the metal.
Typically,  the  traversal  time  is  much  shorter   than   the
relaxation  time  in  the  electron  liquid.  Accordingly, it is
legitimate to separate the tunneling into a single-electron  and
many-electron  parts,  and  to  treat them separately. The first
contribution is described by the transmission coefficient of the
barrier, at small bias being just a constant. We are  interested
here  in  the  multi-electron  effect  that involves motion of a
large number  of  electrons  in  order  to  accomodate  the  new
electron.  At  low  bias  this  collective effect may completely
control the tunneling rate.

Let  us  illustrate  the  effect  of  charge  relaxation  on the
tunneling  rate  by  using  the  example  of  a  two-dimensional
conducting    plane\cite{Spivak}.   Charge   relaxation   in   a
two-dimensional conductor is a classic  electrodynamics  problem
studied  by  Maxwell  who  gave  a solution in terms of a moving
image charge\cite{Maxwell}. In this problem, a point charge  $e$
is  injected into a conducting sheet with conductivity $\sigma$,
and one is interested in the time-dependence of the density  and
potential   of   spreading  charge.  The  solution,  as  Maxwell
formulates it, is that the potential within the sheet  is  given
by  that  of  a  point charge $e$ moving along the normal to the
plane with the velocity $v=2\pi\sigma$. The size of  the  charge
cloud  grows as $r(t)\sim v t$. Let us consider the Coulomb part
of the action for the charge:
  \begin{equation}
{\cal S}(t)\sim \int^t_{t_0}{e^2\over r(t')}dt'=
{e^2\over2\pi\sigma}\ln\left({t\over t_0}\right)
\label{Spivak}
   \end{equation}
In the semiclassical picture the action (\ref{Spivak})  must  be
added   to   the   under-barrier   action.   The  divergence  of
(\ref{Spivak}) at  $t\to\infty$  indicates  that  for  a  charge
spreading  that takes a long time the spreading action dominates
the tunneling rate. We will  see  that  the  time  of  spreading
diverges  at  small  bias,  $t_\ast=e/\sigma V$. From that, near
zero  bias  the  tunneling  acquires  a  suppression  factor  of
$\exp\left(-{1\over\hbar}{\cal  S}(t_\ast)\right)$. The estimate
(\ref{Spivak}) showing that the action diverges  at  small  bias
means  that the semiclassical treatment is meaningful even for a
well conducting metal.  However,  in  the  diffusive  limit  the
estimate  (\ref{Spivak})  does  not  agree with the perturbation
theory. We shall see that the reason is that the  main  part  of
the  action is rather Ohmic than Coulomb, and that after writing
the action properly the semiclassical method completely recovers
the pertubation theory result.

The relevance of the semiclassical picture in this  problem  can
be  justified by a more general argument, not involving specific
features of the charge relaxation  in  two  dimensions.  Let  us
consider a situation when at small bias one electron crosses the
barrier.  Since  the barrier traversal time is much shorter than
the relaxation time in the metal, while the  electron  traverses
the  barrier  other  electrons  practically  do  not  move. Thus
instantly a large electrostatic potential is formed, both due to
the tunneling electron itself, and due  to  the  screening  hole
left  behind. The jump in electrostatic energy by an amount much
bigger than the  bias  $eV$  means  that  right  after  the  one
electron  transfer we find the system in a classically forbidden
state ``under'' the Coulomb  barrier.  In  order  to  accomplish
tunneling,  the  charge  yet has to spread over a large area, so
that the potential of the charge fluctuation  is  reduced  below
$eV$.  If  the  conductivity is finite, the spreading over large
distance  takes  long  time,  and  thus  the  action   of   this
cooperative under-barrier motion is much bigger than $\hbar$.

\section{The action for long-wavelength modes}
   For the electrodynamics problem the action can be written  in
terms  of  charge  and  current  densities $\rho(r,t)$ and ${\bf
j}(r,t)$.  Full  action  would  also   contain   electromagnetic
potentials,  but  in  the  quasistationary  limit, $c\to\infty$,
which we always assume below, the potentials are  ``slaved''  to
charges, and thus can be integrated out. As it was argued above,
the contribution to the action of the spreading charge is mainly
coming   from   long   times  when  the  charge  deviation  from
equilibrium is small. Therefore, we can  expand  the  action  in
powers   of  $\rho(r,t)$  and  ${\bf  j}(r,t)$,  and  keep  only
quadratic terms.  The  action  should  reproduce  the  classical
electrodynamics  equations: the Ohm's law and charge continuity.

Of  course  this requirement is entirely sufficient to determine
the form of the action. However, it is more convenient to  argue
in  the  following  way. We are going to use the action to study
the  dynamics  in  imaginary  time.  Therefore,  the  action  is
precisely   the  one  that  appears  in  the  quantum  partition
function. The latter action expanded up to  quadratic  terms  in
charge  and  current density must yield correct Nyquist spectrum
of current fluctuations in equilibrium:
   \begin{equation}\label{Nyquist}
\langle\!\langle {\bf g}_{\omega,q}^\alpha
{\bf g}_{-\omega,-q}^\beta \rangle\!\rangle =
\sigma_{\alpha\beta}|\omega|+
\sigma_{\alpha\alpha'}D_{\beta\beta'}q_{\alpha'}q_{\beta'}\ .
   \end{equation}
Here
\begin{equation}
{\bf g}={\bf j}+\hat D\nabla \rho
\end{equation}
  is external current and $D_{\alpha\beta}$  is  the  tensor  of
diffusion  constants  related  to the conductivity tensor by the
Einstein's    formula:    $\hat\sigma=e^2\nu\hat    D$,    where
$\nu=dn/d\mu$  is  compressibility. Generally, both $\hat\sigma$
and $\hat D$ are functions of the frequency  and  momentum.  For
simplicity,  we  assume that the temperature is zero and discuss
only a  two  dimensional  metal  with  spatially  isotropic  and
homogeneous       conductivity:       $\sigma_{xx}=\sigma_{yy}$,
$\sigma_{xy}=-\sigma_{yx}$.

The requirement that the action produces correct current
fluctuations is essentially equivalent to the
fluctuation-dissipation theorem. Thus, the form of the action is fixed by
response functions of the system. In imaginary time we get
   \begin{equation}\label{action}
{\cal S}={1\over 2} \int\!\int\! d^4x_1d^4x_2\!
\left[{\bf g}_{1}^{T}
\hat K_{x_1\!-x_2}{\bf g}_{2}+
{\delta_{12}\rho_{1} \rho_{2}\over|r_1-r_2|}\right]
\end{equation}
where $x_{1,2}=(t_{1,2},r_{1,2})$, and $\delta_{12}=\delta(t_1-t_2)$.
The kernel $\hat K_{r,t}$ is  related  to  the  current-current correlator,
  \begin{equation}
(K^{-1}_{\omega,q})_{\alpha\beta}=
\langle\!\langle {\bf g}_{i\omega,q}^\alpha
{\bf g}_{-i\omega,-q}^\beta \rangle\!\rangle
\end{equation}
   given by (\ref{Nyquist}), where $\hat\sigma$ and $\hat D$ are
functions of the Matsubara frequency related with
the real frequency functions by the usual analytic continuation.
We take Coulomb interaction in the second term of the
action (\ref{action}) as non-retarded because we are going to study
systems with relatively low conductivity, and thus slow charge relaxation.

One can also justify the form (\ref{action}) of the action by
looking at various particular examples, like the  ideal
dissipationless liquid of charge, or the phenomenological
dissipative Caldeira-Leggett system with spatially distributed
coupling to a thermal bath. In the first case the Ohmic part of
the action is replaced by the kinetic energy term
  $${\cal S}_{kin}= \int dr dt{{\bf  j}^2\over2e\rho}\ .$$
In  the second  case  one  has
  $${\cal  S}_{CL}=  \int  dr d\omega {{\bf j}_{\omega}(r)\cdot
{\bf j}_{-\omega}(r)\over 4\pi e^2 \eta|\omega|}\ ,$$
   where $\eta$ is the Caldeira-Leggett viscosity. For ${\cal
S}_{kin}$ the electric impedance of the system is imaginary,
while for ${\cal S}_{CL}$ it is real. From these two
``limiting'' cases one can conjecture the more general form
(\ref{action}).

\section{Instanton in imaginary time}
  To evaluate the tunnelling rate, we use the  instanton
method and look for a least action ``bounce''
path in imaginary
time\cite{Langer,Coleman}. Among the bounce paths symmetric in time,
$\rho(r,t)=\rho(r,-t)$, ${\bf j}(r,t)=-{\bf j}(r,-t)$,
we shall find the least action path which will give a
semiclassical estimate of the tunneling rate exponent.

From the  variational  principle  one  can  derive  equation  of
motion.  We  note  that  the  action (\ref{action}) contains the
charge and current densities as {\it independent} variables,  as
Eq.(\ref{action})  was  derived by matching with the equilibrium
fluctuations in the {\it grand canonical ensemble} where  charge
is  not  conserved.  Therefore,  we  have  to  supply the action
(\ref{action})   with   the   charge   continuity    constraint:
$\dot\rho+\nabla\cdot{\bf      j}={\cal      J}(r,t)$,     where
\begin{equation}\label{source}                             {\cal
J}(r,t)=e\delta(r)(\delta(t+\tau)-\delta(t-\tau))\             .
\end{equation}
  This form of  the  charge  source  ${\cal  J}(r,t)$  describes
electron injected the system at $r=0$, $t=-\tau$, and taken back
at $t=\tau$, at the same point. In principle, one could consider
a more general source term ${\cal J}$ of the form
  \mbox{${\cal J}=e(\delta(t+\tau)\delta(r-r_1) - \delta(t-\tau)\delta(r-r_2))$,}
which  corresponds  to  the process of one electron entering and
then leaving the liquid at different points.  However,  in  real
situation the tunnelling occurs preferentially at ``hot spots'',
or  defects, where the tunnelling barrier is low or narrow. Thus
we assume $r_1 = r_2$, which also accounts for the quantum point
contact tunneling experiment.

The charge continuity requirement is incorporated in the
action by adding a Lagrange mutiplier term:
   \begin{equation}
{\cal S}_{total}={\cal S}(\rho,{\bf j})+\phi(r,t)\left(
\dot\rho+\nabla{\bf j}-{\cal J}(r,t)\right)\ ,
\end{equation}
   where $\phi(r,t)$ is an independent variable of the  problem.
For  the  least action path, the variation of ${\cal S}_{total}$
relative to infinitesimal change $\delta\phi$, $\delta\rho$, and
$\delta{\bf  j}$  vanishes.  (Note  that  due  to   the   charge
continuity $\delta\dot\rho+\nabla\cdot\delta{\bf j}=0$.) After
eliminating $\phi$ we get the   standard equations of classical
electrodynamics:
\begin{eqnarray}\label{dynamics}
({\rm i})\qquad & & \dot\rho+\nabla\cdot{\bf j}={\cal J}(r,t)\ ;\nonumber\\
({\rm ii})\qquad & & {\bf j}+D\nabla\rho =\hat\sigma(\omega,q){\bf E}\ ;\\
({\rm iii})\qquad & & {\bf E}(r,t)=-\nabla_r\int dr'{\rho(r',t)U(|r-r'|)}\ .
\nonumber
\end{eqnarray}
The equations describe the system trajectory in imaginary  time,
i.e.,  the  spreading of charge under the Coulomb barrier due to
selfinteraction.

Then one must solve
Eq.(\ref{dynamics})  for  $\rho$ and ${\bf j}$, and to compute
the action (\ref{action}). For a spatially  homogeneous system,
by using Fourier transform, we get
  \begin{eqnarray*}
\rho(\omega,q)&=&{{\cal J}(\omega)\over |\omega|+Dq^2+\sigma_{xx}q^2U_q}\ ,\\
{\bf j}(\omega,q)&=&{-i\hat K^{-1}(\omega,q){\bf q}U_q \rho (\omega,q)}\ ,
\end{eqnarray*}
where $U_q$ is the Coulomb potential formfactor.
Substituted in Eq.(\ref{action}) this yields the action
\begin{equation}\label{Stau}
{\cal S}_0(\tau)=
{1\over2}\sum\limits_{\omega,q}\,
{|{\cal J}(\omega)|^2\over|\omega|+Dq^2}\,
{U_q\over|\omega|+Dq^2+\sigma_{xx}q^2U_q}
\end{equation}
which  depends  on the accomodation time $\tau$ through Fourier
component of the charge source: ${\cal J}(\omega)=2ie\sin\omega\tau$.

Finally,  to  obtain total action of the system we subtract from
the action ${\cal  S}_0(\tau)$  of  spreading  charge  the  term
$2eV\tau$  that  accounts  for  the work done by voltage source:
\mbox{${\cal S}(\tau) = {\cal S}_0(\tau)  -2eV\tau$}.  Thus  the
energy  conservation  at  transferring  one  electron across the
barrier is assured. Then one has to optimize ${\cal S}(\tau)$ in
$\tau$. Optimal $\tau_\ast$ satisfies the relation
  \begin{equation}\label{tau}
\frac{\partial {\cal S}_0(\tau_\ast(V))}{\partial \tau} = 2eV
\end{equation}
  Having found $\tau_\ast$ from
Eq.(\ref{tau}) one gets the
tunneling rate that coinsides with tunneling
conductivity up
to a constant factor:
  \begin{equation}\label{probability}
G(V)=G_0\exp\left[-{1\over\hbar}\left({\cal S}_0(\tau_\ast(V))
-2eV\tau_\ast(V)\right)\right]\ .
\end{equation}
The optimal $\tau_\ast $ can be interpreted as the charge accomodation
time.

Let us point out here that the accuracy of the term $2eV{\tau}$
is determined by the assumption that $\tau_\ast\gg\tau_f$, the
time it takes one electron to traverse the barrier. This
assumption is valid whenever there is an anomaly: if
$\tau_\ast\approx\tau_f$, then ${\cal
S}(\tau_{\ast})\approx\hbar$, and thus there is no tunneling
suppression.

To summarize, the equations (\ref{Stau}), (\ref{tau}), and
(\ref{probability}), taken together, define tunneling
conductivity. Within this quite general framework, one can study
the anomaly in different systems. Let us emphasize that after
having calculated $\tau_\ast(V)$ and ${\cal S}(\tau_\ast)$ it is
essential to check the selfconsistency of the assumption that
$\tau_\ast\gg\tau_f$. For example, this assumption will not be
fulfilled in a {\it clean} metal, i.e., in Fermi liquid without
disorder ($D>1$). The reason is that the conductivity of an
ideal conductor is $\sigma(\omega)=ine^2/m\omega$. In this case
Eq.(\ref{Stau}) gives ${\cal S}_0(\tau)\approx\hbar$ at any
$\tau\gg\tau_f$, and henceforth $\tau_\ast\simeq\tau_f$. This
indicates absence of the anomaly in a clean metal, the result
familiar from the Fermi liquid picture. On the contrary, for a
one-dimensional metal ${\cal S}_0(\tau)\sim \ln\tau/\tau_f$,
which leads to the power-law anomaly known from the Luttinger
liquid theory.

\section{Comparison with diffusive anomaly}
For a two-dimensional metal with elastic scattering time $\tau_0$ and
non-screened Coulomb interaction we set
$U_q = 2\pi/|q|$ and
$\sigma_{xx}=\sigma$,  constant at $|\omega|, v_F|q|\le 1/\tau_0$. Then
Eq.(\ref{Stau}) gives
\begin{equation}\label{AAL}
{\cal S}(\tau) =\,
{e^2\over 8\pi^2\sigma}\,
\ln \left({\tau\over\tau_0}\right)\,
\ln \left(\tau \tau_0 \sigma^2 (\nu e^2)^2 \right)\ .
\end{equation}
(Here $\nu$ is compressibility.) From Eq.(\ref{tau}),
 \begin{equation}\label{AAL_tau}
\tau_\ast = \frac{e}{4\pi^2 V\sigma}\,
\ln(\hbar\sigma\nu e/V)\ .
\end{equation}
The theory is selfconsistent in the hydrodynamic limit,
$\tau_\ast\ge\tau_0$, i.e., at $eV\le e^2/\sigma\tau_0$.
Then the least action is
\begin{equation}
\label{DoubleLog}
{\cal S}(V) = \frac{e^2}{8\pi^2\sigma}\,
\ln\left(\frac{e}{4\pi^2\sigma V\tau_0}\right)\,
\ln\left(\frac{e\tau_0\sigma(\nu e^2)^2}
{4\pi^2 V}\right)
\end{equation}
   It is interesting to compare this result with the identical
double-log dependence derived by Altshuler, Aronov, and Lee in a
different context\cite{AAL}. They calculated perturbatively the
correction to the tunneling density of states
$\delta\nu(\epsilon)$ with the assumption that it is small,
$|\delta\nu|\ll\nu_0$, which is the case only for a
weak disorder. It was found that
$\delta\nu(\epsilon)=-\hbar^{-1}\nu_0{\cal S}(V=\epsilon/e)$,
where ${\cal S}(V)$ is given by (\ref{DoubleLog}). The main
difference is that our double-log has to be exponentiated to get
the tunnelling density of states, while in \cite{AAL} the
double-log itself appears as a correction to the density of
states. In the range of the perturbation theory validity the two
results agree. From that point of view, our calculation provides
description of the diffusive anomaly at low bias, where the
perturbation theory diverges.

\section{Screening by electrodes}
In a real experiment the charge tunnels between two
electrodes, and there are separate contributions to the action due to
the relaxation of the electron and hole charges on both sides of
the barrier. If the electrodes are close, the
charges partially screen the field of each other, which
makes their spreading correlated. In this case the least action
is smaller than the sum of independent contributions of the
electrodes, and thus the anomaly is weakened. For a two
dimensional system, quantitatively, the effect will be that
the log-divergence of the integral over $q$ in Eq.(\ref{Stau})
will be cut at $q\simeq a^{-1}$,
where $a$ is the
distance between the electrodes. As a result,
the $V-$dependence of the second log in Eq.(\ref{DoubleLog})
saturates at $eV\simeq V_0=\hbar\sigma/a$.

This ``excitonic'' correlation effect can be treated
straightforwardly by writing the action (\ref{action}) for each
electrode separately, together with the term describing
interaction across the barrier.
Let us consider an example of two parallel planes with different
conductivities and diffusion constants, $\sigma_1$, $\sigma_2$,
$D_1$, and $D_2$. It is straightforward to generalize the action
and to find the instanton. The least action is
   \begin{equation}\label{TwoPlanes}
{\cal S}_0(\tau)=\frac{1}{2}
\sum\limits_{\omega,q}\,
|{\cal J}(\omega)|^2
\left[\matrix{1\cr -1\cr}\right]^\top
\left(|\omega|+{\bf D}q^2\right)^{-1}
{\bf U}\left(|\omega|+{\bf D}q^2+{\bf\Sigma}q^2{\bf U}\right)^{-1}
\left[\matrix{1\cr -1\cr}\right]\ ,
\end{equation}
  where we use matrix notation:
\begin{eqnarray}\label{matrices}
{\bf D}=\left(\matrix{\hat D_1&0\cr0&\hat D_2}\right)\,,&\ &
{\bf\Sigma}=\left(\matrix{\hat\sigma_1&0\cr0&\hat\sigma_2}\right)\,,\
{\bf U}=\left(\matrix{U_q&V_q\cr V_q&U_q}\right),\nonumber\\
U_q=\int e^{iqr}{d^2r\over |r|}&\ &,\
V_q=\int{e^{iqr}\over\sqrt{r^2+a^2}}d^2r\ .
\end{eqnarray}
  The  rows  and  columns  of  the   matrices   (\ref{matrices})
correspond to the two planes.

Let us assume $e^2\nu_{1,2}a\gg1$, which is the case in almost all
experiments.  Then, by  carrying out matrix inversion and integration we get
  \begin{equation}\label{otvet}
{\cal  S}_0(\tau)=\alpha\ln\left({\tau\over\tau_0}\right)\ {\rm at}\
\tau\gg\hbar/ eV_0\ .
\end{equation}
   Here
  \begin{equation}\label{alpha}
\alpha={e^2\over4\pi^2}
\left[{1\over\sigma_{1,xx}}\,\ln{4\pi\sigma a\over D_2}
+ {1\over\sigma_{2,xx}}\,\ln{4\pi\sigma a\over D_1}\right]\ ,
\end{equation}
  where
$\sigma=\sigma_{1,xx}\sigma_{2,xx}/(\sigma_{1,xx}+\sigma_{2,xx})$.
If the two planes are identical,
\begin{equation}\label{IdenticalSigma}
\alpha={e^2\over2\pi^2\sigma_{xx}}\ln2\pi e^2\nu a\ .
\end{equation}
Thus at $V<V_0$ the $I-V$ curve becomes the power law $I\sim V^{\alpha+1}$.
The tunneling suppression in this case is weaker than for the
non-screened interaction.

Like other interaction effects, the anomaly is enhanced if the system
dimension is lowered. Let us consider tunneling into a disordered wire of
thickness $d$ from a well conducting electrode, two- or
three-dimensional, parallel to the wire at
a distance $a\gg d$ away from it. In this case the main
contribution to the action is due to charge relaxation in the wire. We still
can use Eq.(\ref{Stau}), where $\sum\limits_q...=\int{dq\over2\pi}...$
and the two-dimensional $\sigma_{xx}$ is replaced by $A\sigma$, where $A$ is
the wire crossection and $\sigma$ is the three-dimensional conductivity.
The Coulomb potential formfactor at small $k\ll a^{-1}$, using
electrostatic image in the electrode, is found to be
$U_0=2\ln a/d$. By doing the sum in (\ref{Stau}) over $k$ and
$\omega$ we
get
  \begin{equation}\label{Swire}
S_0(\tau)=B\sqrt{\tau}\ ,\ \ B={U_0 e^2\over \sqrt{\pi D}+
\sqrt{\pi(D+\sigma U_0A)}}\ .
\end{equation}
  Then Eq.(\ref{tau}) gives optimal $\tau_\ast(V)=B^2/16e^2V^2$,
and thus the conductivity
  \begin{equation}\label{AnomalyWire}
G(V)\sim\exp\left(-{B^2\over 8\hbar eV}\right)\ .
\end{equation}
  The ``field$+$charge'' diffusion constant $\sigma U_0 A$ that
appears in the 1D plasmon dispersion relation $\omega(k)=-i\sigma U_0
A k^2$ is typically much bigger than $D$.
Then the spreading of charge is effectively one dimensional at
$V\le V_1=e^2U_0/8a$. For a thick wire Eq.(\ref{AnomalyWire}) holds
in a wide voltage range
$V_1\ge V\ge V_{diff}$
where one can ignore localization corrections to $\sigma$. By the Thouless
criterion, the voltage $V_{diff}$ above which the diffusive treatment is
valid is estimated by comparing the tunneling time $\tau_\ast(V)$ with
the inverse level
spacing in the region over which the charge spreads,
$\Delta^{-1}=\nu A\sqrt{\sigma U_0 A\tau_\ast(V)}$.
This gives
   \begin{equation}\label{dif_limit}
V_{diff}=\left(4\sqrt{\pi}\hbar\nu^2 e D A^2\right)^{-1}\ .
\end{equation}
  On the time scale $\tau_\ast(V_{diff})$, the transport is diffusive
if the wire length $L$ is much shorter than $L_{diff}=\sqrt{\sigma U_0
A \tau_\ast(V_{diff})}=\hbar \nu A^2 \sigma U_0$.

\section{Tunneling in a finite system}
   It is of interest to consider the tunneling problem for a
system of small size in which the time of charge spreading is
limited by the dimension. Most important realization is
tunneling in small Coulomb blockade devices, like quantum dots
or very small capacitors. For such systems, at sufficiently low
bias, the dynamics of charge spreading is unimportant and the
effect of charge relaxation can be taken into account by
modelling the conducting part of the system by a single
effective resistor.

In this limit, the problem can be treated by the recently
developed theory of the ``environment effect'' on the tunneling
conductivity\cite{Nazarov,GirvinGlazman,Devoret}. In the latter
the environment of a tunnel junction is replaced by an effective
electric circuit that causes voltage fluctuations across the
junction. Assuming that the fluctuation spectral density is
equilibrium, one calculates the conductivity suppression factor
by doing gaussian average of the tunneling current, in a way
similar to the Debye-Waller theory of the fluctuation effect on
the structure factor of crystals\cite{Nazarov,Devoret}.
Alternatively, one can calculate the tunneling transition rate
that includes the shake up effect of the plasmon modes in the
environment\cite{GirvinGlazman}. The results of
Refs.\cite{Nazarov,GirvinGlazman,Devoret} can be derived by our
method if one takes the zero wavelength limit of the action
(\ref{action}) by keeping only the largest scale electromagentic
mode for each element of the circuit. (For example, in a
resistor one takes uniform current distribution and zero charge
density, or in a capacitor one assumes uniform charge
distribution over the capacitor plates and no current.) Then the
instanton calculation will lead to the results identical to
those of Refs.\cite{Nazarov,GirvinGlazman,Devoret}.
This zero wavelength limit describes tunneling at very low
voltage, when during the under-barrier motion the tunneling charge
spreads over a scale comparable to
the size of the system.

Therefore, there are several regimes corresponding to different
values of the bias. Let us list them here for the problem of
tunneling between two identical conducting sheets $L\times L$ of
conductivity $\sigma$, separated by
the barrier of thickness $a$.
  \begin{itemize}
\item Effectively infinite system: $e^2/a>eV>e^2a/L^2$
\begin{equation}
G=dI/dV\sim V^{\alpha},\ \ \alpha={e^2\ln(e^2\nu a)  \over4\pi h\sigma}\ ;
\end{equation}
\item Finite system characterized by impedance of environment:
$e^2a/L^2>eV>\hbar D/L^2$
\begin{equation}
G\sim V^{\alpha},\ \ \alpha={e^2\over4\pi h\sigma}\ln L/d\ ;
\end{equation}
\item Ohm's law restored at $eV<\hbar D/L^2$: G={\rm const}
   \end{itemize}
In the expression for $\alpha$ given  for  the  infinite  system
it  is  assumed that $e^2\nu a\gg1$. The result for finite
systems agrees  with  Refs.\cite{Nazarov,GirvinGlazman,Devoret}.
The  third  case  describes the limit of very low bias, when the
system is characterized by an overall transmission  coefficient,
which  is equivalent to connecting in series the barrier and the
metal effective resistors. The three cases describe  the  change
of   the   conductivity  over  the  whole  voltage  range,  from
relatively high voltage where perturbation theory works, through
the region of low voltage where the anomaly disappears.

\section{2D electron gas in magnetic field}
It is straightforward to incorporate magnetic field in the
theory by substituting $\sigma_{xx} =
\sigma_{xx}(B)$ in the action (\ref{AAL}). (Note that $\sigma_{xy}$
does not enter.)
As magnetic field increases,
the conductivity drops, and at certain
field it reaches the quantum limit $\sigma_q=e^2/\hbar$.
In this field range the
prefactor $\alpha$ in Eq.(\ref{otvet}) becomes of the order of one, and the
anomaly in the conductivity changes from weak to strong. The
threshold conductivity, according to Eq.(\ref{AAL}), is
  \begin{equation}\label{threshold}
\sigma_c={1\over4\pi^2}{e^2\over\hbar} \ln2\pi e^2\nu a
\end{equation}
  A transition like that was observed by Ashoori et al.\cite{Ashoori} in
the tunneling current from a 3D metal into a 2D electron gas. In
this experiment, the ohmic conductance was measured as function of
temperature, which corresponds to our zero temperature
non-linear current taken at $V\simeq k_B T/e$. The 2D gas was
relatively clean with the zero field mobility corresponding to the elastic
scattering time $\tau_0\simeq 4\cdot 10^{-12}\ s$. The Fermi
energy calculated from the electron density was $E_F\simeq 10\
mV$. By using the result (\ref{threshold}) together with the
Drude-Lorentz model,
$\sigma_{xx}(B)=ne^2/m\tau_0\omega^2_c$ at $\tau_0\omega_c\gg 1$,
one finds that the anomaly hardening transition
corresponds to the cyclotron frequency $\omega^{\ast}_c=(8\pi
E_F/\hbar\tau_0)^{1/2} \simeq 8.0\ mV$. In terms of the field
intensity this is approximately $4.6\ Tesla$ which is quite
close to the transition field reported in Ref\cite{Ashoori}.

It is interesting to note that in a weakly disordered metal with
$E_F\tau_0\gg\hbar$ the threshold field is small:
$\hbar\omega^{\ast}_0\ll E_F$. This means that the transition
occurs well below the field where the Quantum Hall state is
formed. Therefore, our estimate of $\omega^{\ast}_0$ based on
the ``bare'' Drude-Lorentz conductivity is
meaningful and legitimate. On the other hand, to find the
current at very low $V$ one would have to use the conductivity
renormalized by localization and interaction effects.

Finally,  let  us mention a relation with the work
by Halperin, He, and Platzman\cite{HeEtAl} that deals with the
anomaly in the $\nu=1/2$ Quantum Hall state. In this work,  the
problem   was   treated  by  summing  linked  cluster  terms  of
perturbation theory, with the density responce function borrowed
from the Chern-Simons Fermi liquid theory\cite{HalperinLeeRead}.
The anomaly was found to have the form:
  \begin{equation}\label{nu=1/2}
G(V)\sim\exp\left(-V_0/V\right)\ , \ \ V_0=4\pi{e\over\epsilon}\sqrt{\pi n}\ ,
\end{equation}
   where $V\ll V_0$, and $n$ is density. It  is  interesting  to
see how this result can be derived from the effective action. It
has  been  shown\cite{HalperinLeeRead}  that conductivity of the
$\nu=1/2$ state has strong spatial dispersion:  $\sigma_k=A|k|$,
$A=e^2/16\pi\epsilon\sqrt{\pi  n}$.  If this form is inserted in
the action (\ref{Stau}), one  gets  $S(\tau)=\pi\sqrt{2\tau/A}$,
which leads to the tunneling rate (\ref{nu=1/2}).

\section{Conclusion}
The essential feature of  the  presented  approach  is  that  it
relates  the  tunneling  anomaly  with exact conductivity of the
system without calculating it. This  would  allow  a  comparison
with  experiment  in  the  situations where there is no accepted
model for conductivity. For example, the  tunneling  current  as
function  of  voltage can be taken from experiment, and directly
used   to   find   ${\cal   S}(\tau)$   by   inverse    Legendre
transformation.  Thus  obtained function ${\cal S}(\tau)$ can be
analyzed to extract the conductivity  frequency  and  wavevector
dependence.

To  summarize,  we argued that the theory of the Coulomb anomaly
in  the  regime  of   strong   suppression   of   tunneling   is
semiclassical. The underlying reason is that the transfer of one
electron  across the barrier is controlled by cooperative motion
of many other electrons.  We  treat  this  motion  as  classical
electrodynamics  in  imaginary  time,  write down the action and
find the least action  instanton  trajectory.  The  results  are
compared to several known perturbation theories and an agreement
is  found.  The theory is used to interpret experiments
on the coulomb gap induced by magnetic field.

\begin{center}
\small{\bf
BLOCKAGE COULOMBIEN DE L'EFFET TUNNEL:\\ DE L'ANOMALIE \` A ZERO VOLTAGE
AU GAP COULOMBIEN}
\bigskip
\end{center}
\bigskip

{\small
Nous pr\'esentons une approche de la th\'eorie effective pour
le probl\`eme du blockage coulombien dans l'effet tunnel. La
m\'ethode est appliqu\' ee au probl\`eme de la limite de fort
couplage proche du voltage zero, ou la th\'eorie de la
perturbation diverge. Par un argument semiclassique, nous
obtenons une th\'eorie electrodynamique en temps imaginaire, et
nous exprimons l'anomalie en terme de la conductivit\'e exacte
$\sigma(\omega, q)$ du syst\`eme et de l'interaction exacte.
La validit\'e de notre approche est test\'e par comparaison
avec le calcul  connu de l'anomalie diffusive obtenu en
th\'eorie de la perturbation.
   Nous  utilisons  aussi  cette  methode  pour   l'\'etude   de
l'accroissement  de  l'anomalie  du  \`a  un champs magn\'etique
externe, et \`a l'\'ecrantage des \'electrodes.}
  \bigskip

\end{document}